\documentclass[
 amsmath,amssymb,aps,
 reprint,%
]{revtex4-2}

\usepackage{graphicx}
\usepackage{dcolumn}
\usepackage{bm}

\usepackage[mathlines]{lineno}

\begin{document}

\preprint{AAPM/123-QED}

\title{A Comparative Study of Shadows of Magnetized and Non-Magnetized Singularities}

\author{Amee Bhatt}
 \affiliation{Department of Physical Sciences, PDPIAS, CHARUSAT University}
 \email{ameeb9@gmail.com}

\date{\today}

\begin{abstract}
The recent observations of the galactic center of the M87 galaxy have made the field of observing black holes and calculating its shadow much more intriguing. Approaching the question of calculating shadows, many approximations are made in order to simplify the equations which makes the considered case less realistic. Understanding the shadow of different singularities under the influence of magnetic field is of more importance astrophysically as the accreting matter around the singularity would generate electromagnetic fields as it would be in plasma state due to the high tidal effects. Here, we use Ernst technique to immerse spacetimes in uniform, sourceless and asymptotic magnetic field. Later, we compare the effective potential of null geodesics in magnetized and non-magnetized cases. This study would be helpful in understanding the M87 shadow and the forthcoming image of shadow of SagA*.
\end{abstract}

\maketitle
\section{\label{sec:level1}Introduction}
The EHT collabration, by releasing the first image of black hole at the center of the M87 galactic center have opened in new horizons in the field of astrophysics and black hole physics\cite{akiyama2019first}. Considering the astronomical distances between any black holes and us, every observation would always be indirect, as we would only be able to see the shadow due to the strong gravitational lensing effect. There are different types of physical conditions in the vicinity of a black hole. 
Including magnetic field in the calculations of the shadow would generate results different than the non-magnetized case. Theoretically, there are many possible solutions to Einstein's field equation other than black hole. A naked singularity in one such example. The null geodesics/light can escape the singularity and would be observable to a distant observer if certain conditions during the gravitational collapse are fulfilled. The singularity could thus be locally or globally visible which depends on the time of formation of trapped surfaces and apparent horizon.\cite{wald2010general} Here, in II we discuss the different conditions for shadow formation, in section III, the Melvin universe and Ernst's technique is summarized, section IV the null-like naked singularity is immersed in the Melvin universe, The section V we discuss and compare the effective potential plots of magnetized and non-magnetized spacetimes of schwarzschild black holes and null naked singularity.

\section{Shadow formation Conditions}
“In the asymptotic observer’s sky, the image of a compact object may consist of a central dark region where at the boundary of the dark region, the intensity of light is maximum and decreases gradually as radial distance from the central compact object increases. The central dark region in the distant observer’s sky, is known as the Shadow of the compact object” \cite{kaur2021comparing}. \cite{}
 \begin{figure}
     \centering
      \includegraphics[width=90mm,scale=0.2]{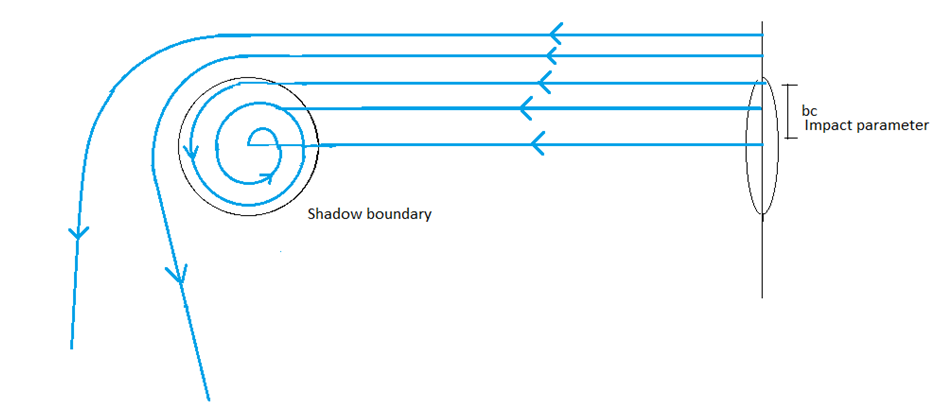}
 \caption{Shadow ray diagram}
     \label{fig:my_label}
 \end{figure}
The structure with central dark spot and ring of light with high luminosity formed due to the presence of a highly compact object which distorts spacetime, at the center with light.
\\For the shadow formation, the light coming from a distant object encounters the compact object and behaves in accordance with the metric formed due to the influence of the compact object to give shadow to an asymptotic observer.
\\As shown in figure (1) the light coming from the Impact parameter region would be totally lost to the distant observer as it would directly plunge in the central compact object.
\\Formation of a shadow depends upon the metric of the spacetime around the compact object, hence, for different compact objects the physical signature of the shadow is different for an asymptotic observer.
\\Effective potential vs radial distance from the central compact object graph describes the possibility of formation of shadow. The existence of upper bound of $V_{eff}$ of null geodesics is the necessary condition for shadow formation.
Shadow can exist even without the photon sphere. For the photon sphere to exist second order derivative of $V_{eff}$ with respect to radial distance should be greater than zero.
\\The derivation of the formula of $V_{eff}$ can be given as follows:
Consider a spherically symmetric static spacetime. Its line element can be given by: 
\\ \begin{equation}
    ds^2= -g_{tt}(r)dt^2 + g_{rr}(r)dr^2 + S(r)(g_{\theta\theta}d\theta^2 + g_{\phi\phi} d\phi^2)
\end{equation}
where $g_{tt}, g_{rr}$ are the functions of r only, and the azimuthal part of the spacetime shows the spherical symmetry. For a particle which is freely falling in this type of spacetime, the angular momentum (h) and the energy $(\gamma)$ per unit of particle’s rest mass are always conserved, where the angular momentum and energy conservation are the direct consequence of spherical and temporal symmetry of the above spacetime.\cite{bambhaniya2019timelike}
\\The conserved angular momentum (L) and the energy (E) per unit of particle’s rest mass can be written as:\cite{bambhaniya2019timelike}
 \begin{equation}
    L = r^2 \frac{d\phi}{d\tau} ; \hspace{1cm}E = g_{tt} \frac{dt}{d\tau} ; \hspace{1cm}\tau = proper time;
\end{equation}
The freely falling particles follow timelike geodesic for which $\nu^u \nu_{u}=-1$
where $\nu^u$=four velocity of free falling particle. From this normalization of four-velocity we can write an effective potential which plays a crucial role on particles’ trajectories in a spacetime.\cite{kaur2021comparing} The effective potential can be given by:
\begin{equation}
    V_{eff} = \frac{1}{2} [g_{tt}(r)(1+ \frac{L^2}{r^2})-1]
\end{equation}
 For the calculations of impact parameter:
Consider angular momentum and energy per unit rest mass in terms of affine parameter $\lambda$
\begin{eqnarray}
    L = S(r)g_{\phi\phi}\frac{d\phi}{d\lambda}  \hspace{1cm}\\
    E = g_{tt}\frac{dt}{d\lambda}
 \\\frac{1}{b^2} = \frac{g_{tt}g_{rr}}{h^2}(\frac{dr}{d\lambda})^2 + V_{eff}
\\\frac{1}{b^2} = \frac{E^2}{L^2}
\end{eqnarray}
Here; b is the impact parameter as mentioned in the figure (1).
\\In a similar way, the effective potential for Schwarzschild spacetime can also be considered and the graph of effective potential vs radial distance can be plotted and the lensing effect of the shadow can be observed.\cite{kaur2021comparing}
\\Mathematically, the condition for the existence of the photon sphere is the second order derivative of the effective potential with respect to the radial distance should be less than zero. Hence,
\begin{eqnarray}
\frac{d^2 V_{eff}}{d^2 r}\\
\frac{dV_{eff}}{dr} = 0
\end{eqnarray}

\begin{figure}
    \includegraphics[width=80mm,scale=0.2]{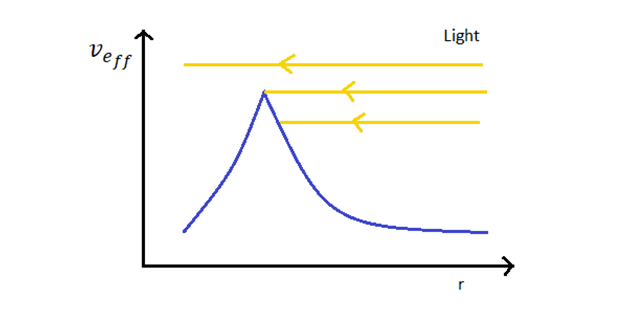}
    \caption{Effective potential with upper bound}
    \label{fig:2}
\end{figure}

\begin{figure}
      \centering
      \includegraphics[width=80mm,scale=0.2]{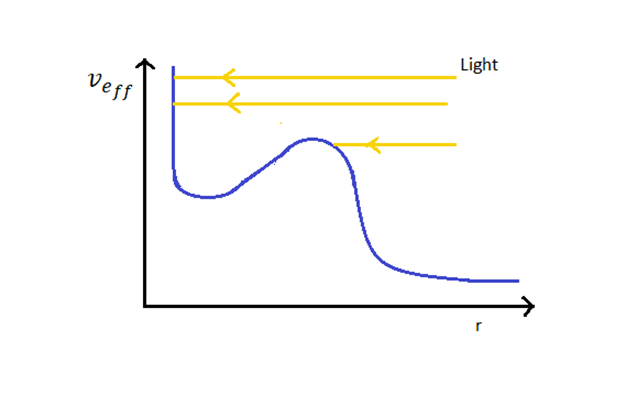}
      \caption{Effective potential without upper bound}
      \label{fig:3}
  \end{figure}

Following are some different cases of the shadow formation:
\\Shadow without photon sphere:Figure\ref{fig:2}
 Here, \(\frac{dV_{eff}}{dr} = 0\) doesn't hold and the condition \( \frac{d^2 V_{eff}}{d^2 r} < 0\) holds so the shadow exist without photon sphere.
 \\Photon sphere without shadow:Figure\ref{fig:3}
  Here, \(\frac{dV_{eff}}{dr} = 0\) and \( \frac{d^2 V_{eff}}{d^2 r} < 0\) also exist which confirms the existence of the photon sphere but there is no existence of upper bound which means the shadow doesn't exist.

\section{Melvin Universe and Ernst's Technique}
In 1952 W.B. Bonner gave solutions to the empty space having cylindrical symmetry containing electromagnetic field and gave its physical interpretations.\cite{rizwana2015black}. This was later rediscovered by M.A. Melvin in 1964.\cite{melvin1964pure}. 
\\Later in 1974, Robert M. Wald \cite{wald1974black}\cite{}derived the solutions for the electromagnetic field generated due to a static and axially symmetric black hole placed in a uniform magnetic field aligned with the axis of the black hole itself. They considered the Killing vectors in a Minkowski type spacetime are equivalent to the vector potential for the Maxwell test field. This paper essentially examined the changes in the magnetic field in the vicinity of said black hole after its placement in the field. The approach to magnetizing the spacetime near a black hole eventually leads to the black hole accreting charge until a certain value and the fields consequently will not remain sourceless. \cite{wald1974black}
\\After all these major strides in the magnetizing the black hole solutions to the combined Einstein-Maxwell, many papers and articles were published generalizing the Ernst technique for immersing black holes in a magnetic universe.

The metric describing the Melvin universe can be written as;\cite{ernst1968new}\cite{ernst1976black}
\begin{equation}
    ds^2 = (1+\frac{1}{4}B^2\rho^2)^2 (-dt^2+d\rho^2+dz^2)+(1+\frac{1}{4}B^2\rho^2)^{-2} \rho^2d\phi^2
\end{equation}
for $t,z\in(-\infty,+\infty)$, $\rho\in[0,\infty)$, $\phi\in[0,2\pi)$. The electromagnetic field can be described by the Maxwell's tensor:
$$F=e^{-i\psi}B(dz\wedge dt)$$
where $\psi$ is a real parameter for the duality rotation. When $\psi=0$ the maxwell tensor in the above equation describes an electric field pointing along the z-direction. Now, $\psi=\frac{\pi}{2}$ the maxwell's tensor would become
\begin{equation}
    F=B(1+\frac{1}{4}B^2\rho^2)^{-2} \rho d\rho \wedge d\phi
\end{equation}
representing only magnetic field aligned along the
z-direction.
\\The Melvin universe describes a spacetime which is static, cylindrically symmetric and where the axial magnetic field is aligned along the z-axis whose magnitude is governed by the parameter B. This solution essentially depicts a universe filled with parallel magnetic or electric fluxes bound together by its own gravitational pull.
\\If the parameter B is taken to be zero, then the above solution represents the Minkowski metric in the cylindrical coordinate symmetry.
The line element of the Minkowski spacetime i.e. asymptotically flat spacetime can be written as:\cite{rizwana2015black}
\begin{equation}
    ds^2=-dt^2+dz^2+d\rho^2+\rho^2d\phi
\end{equation}
similarly, the general form of stationary and axially symmetric line element can be written as:
\begin{equation}
    ds^2=f^{-1}[-2P^{-1}d\zeta d\zeta^{*}+\rho^2dt^2]-f(d\phi-\omega dt)^2
\end{equation}
On comparing equations (3.37) and (3.38), following results are obtained:
\begin{equation}
    f=-\rho^2,  \omega=0;  P=\rho^{-1};  d\zeta=\dfrac{(dz+\iota d\rho)}{\sqrt{2}}
\end{equation}

The complex gravitational potential can be defiled by:
$$\epsilon=f-|\Phi|^2+\iota\phi$$
here $\Phi$ is the complex electromagnetic potential. It's real part represents the electrostatic potential and imaginary part represents the magnetostatic potentials.
In the above equation, $\phi$ represents the twist potential. If the symbol $\nabla$ is given by:
$$\nabla=r\frac{\partial}{\partial r}+\iota\frac{\partial}{\partial\theta}$$
then the twist potential can be obtained from the following equation:
$$-\rho^{-1}f^2\nabla\omega=\iota\nabla\phi+\Phi^{*}\nabla\Phi-\Phi\nabla\Phi^{*}$$

Let, $E_r, E_\theta, H_r, H_\theta$ be the radial and the angular components of the electric and magnetic field respectively, the complex electromagnetic potential can be written in the terms of:
\begin{equation}
    H_r +\iota E_r=P\frac{\partial\Phi}{\partial\phi}; H_\theta+\iota E_\theta=-P\frac{\partial\phi}{\partial r}
\end{equation}

As, there is no electromagnetic field initially, the complex gravitational and electromagnetic twist potentials can be given by:
$$\epsilon=f=-\rho^2;\hspace{1mm} \Phi=0;\hspace{1mm} \Phi=0$$

Hereby, using \textbf{Harrison's transformations} the new functions are as follows:
\begin{equation}
    \Lambda=1+B\Phi-\frac{1}{4}B^2\epsilon
\end{equation}
\begin{equation}
    \epsilon'=\Lambda^{-1}\epsilon
\end{equation}
\begin{equation}
    \Phi'=\Lambda^{-1}\left(\Phi-\frac{1}{2}B\epsilon\right)
\end{equation}
Here the functions f and $\omega$ are transformed as:
\begin{equation}
    f'=Re\epsilon'+|\Phi'|^2= |\Lambda|^{-2}f
\end{equation}
\begin{equation}
    \nabla\omega'=|\Lambda|^2\nabla\omega+\rho f^{-1}(\Lambda^{*}\nabla\Lambda-\Lambda\nabla\Lambda^*)
\end{equation}
The operator $\nabla$ is different for different cases and the functions P and $\rho$ remains unmodified.
Now;\begin{equation}
    \Lambda=1+\frac{1}{4}\rho^2B^2
\end{equation}
As, $\Lambda$ is real the equation of $\omega'$ simply reduces to $\Lambda'=|\Lambda|^2\nabla\omega$ but $\omega=0$ so, $\omega'=0$

Finally, using transformed functions and putting in the equation (3.38) the transformed line element is obtained:
\begin{equation}
    ds^2=\left(1+\frac{1}{4}\rho^2B^2\right)[-dt^2+dz^2+d\rho^2]+\left(1+\frac{1}{4}\rho^2B^2\right)^{-2}d\phi^2
\end{equation}
The transformed electromagnetic potential is given by:
\begin{equation}
    \Phi'=\frac{1}{2}\Lambda^{-1}B\rho^2
\end{equation}
Also, the magnetic field components are given by:
$$H_z=\Lambda^{-1}B^2;  H_{\rho}=0=H_{\phi}$$
\\Similar calculations are done for Schwarzschild, Kerr, and RN black holes as well in Ref\cite{rizwana2015black}.
\\The null-like naked singularity is a non-rotating, charge-less and static solution to the Einstein's field equations which is much more analogous to the Schwarzschild black hole and hence the magnetized schwarzschild solutions are key in understanding the magnetized null-like naked singularity.
The magnetized SCH line element is given as:\cite{rizwana2015black}
\begin{equation}
    ds^2= |\Lambda|^2\left[-\left(1-\frac{2M}{r}\right)dt^2+\frac{dr^2}{(1-\frac{2M}{r})}\right]+|\Lambda|^{-2}r^2sin^2\theta d\phi^2
\end{equation}
Where:
\begin{equation}
    \Lambda=1+B\Phi-\frac{1}{4}B^2\epsilon=1+\frac{1}{4}B^2r^2sin^\theta
\end{equation}

\section{Magnetizing Null-like Naked Singularity}
The null-like naked singularity metric can be given by the following line element:\cite{joshi2020shadow}
\begin{equation}
    ds^2= -\frac{dt^2}{(1+\frac{M}{r})^2}+\left(1+\frac{M}{r}\right)^2dr^2+r^2d\theta^2+r^2sin^\theta d\phi^2 
\end{equation}
On simplifying the above equation we get:
\begin{equation}
    ds^2=-\frac{r^2dt^2}{(r+M)^2}+\frac{(r+M)^2}{r^2}dr^2+r^2d\theta^2+r^2sin^2\theta d\phi^2
\end{equation}
Comparing the above equation with
\begin{equation}
    ds^2=f^{-1}[-2P^{-1}d\zeta d\zeta^{*}+\rho^2dt^2]-f(d\phi-\omega dt)^2
\end{equation}
the following parameters are obtained:
\begin{equation*}
    f=-r^2sin^2\theta;\hspace{1mm} \omega=0;\hspace{1mm} P=(rsin\theta)^{-1}
\end{equation*}
\begin{equation}
    \rho=(r+M)^{-1}r^2sin^2\theta; \hspace{1mm} d\zeta=\frac{1}{\sqrt{2}}\left(\frac{(r+M)}{r}dr+\iota d\theta\right)
\end{equation}
Now using equation (21);
\begin{equation}
    \Lambda=1+B\Phi-\frac{1}{4}B^2\epsilon
\end{equation}
\begin{equation}
    \Lambda=1+\frac{1}{4}B^2r^4sin^2\theta
\end{equation}
To check weather the equations (29) holds true for the null-like naked singularity, it can be crosschecked by putting these values in (28):
\begin{multline}
 ds^2=(-r^2sin^2\theta)^{-1}[-2((rsin\theta)^{-1})^{-1}\\(\frac{1}{\sqrt{2}}(\frac{(r+M)}{r}dr))^2- d\theta^2)+((r+M)^{-1}r^2sin^2\theta)^2dt^2 ]-\\(-r^2sin^2\theta)(d\phi+\omega dt)^2
\end{multline}

The null-like naked singularity metric can be easily obtained from the above equation
The transformed functions f' and $\omega'$ are obtained according to the equations in the previous chapter:
\begin{equation}
     f'=|\Lambda|^-2f
     =-\frac{r^2sin^2\theta}{(1+\frac{1}{4}B^2r^4sin^2\theta)}
    \end{equation}
\begin{equation}
    \omega'=0
    \end{equation}
Here, P and $\rho$ remains unmodified.
The final transformed line element of the magnetized null-like naked singularity can be written as;
\begin{equation}
    ds^2=|\Lambda|^2\left[-\frac{dt^2}{(1+\frac{M}{r})^2}+(1+\frac{M}{r})^2dr^2+r^2d\theta^2\right]+|\Lambda|^{-2}r^2sin^2\theta
    \end{equation}
The combined equations (31) and (35) are crucial for further investigations.

\section{Comparisons of effective potential}
The effective potential for static, spherically symmetric and axial spacetime can be given by:
\begin{equation}
    V_{eff}=\frac{g_{tt}}{r^2}
\end{equation}
The effective potential of magnetized null naked singularity is given as:
\begin{equation}
    V_{eff}=\left|1-\frac{B^2r^2sin^2\theta}{4(r+M)^2}\right|^{-2} (r+M)^{-2}
\end{equation}
Now, the graph of $V_{eff}$ vs radial distance r can be plotted to examine the shadow formation conditions as explained in section-II.
Firstly, we plot the graph of the non-magnetized null-like naked singularity using equation (35) as a reference. The effective potential for this metric is given as:
\begin{equation}
    V_{eff}=\frac{g_{tt}}{r^2}
\end{equation}
\begin{equation}
    V_{eff}=\frac{1}{(r+M)^2}
\end{equation}
\begin{figure}
      \includegraphics[width=95mm,scale=0.2]{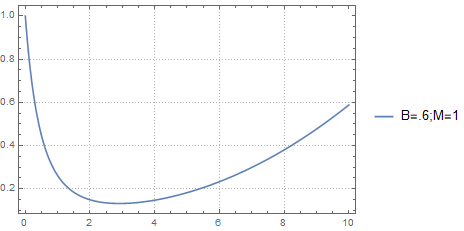}
      \caption{$V_{eff} \longrightarrow r$ of magnetized naked singularity (M=1; B=0.6; $r=0\longrightarrow 10$)}
      \label{fig:4}
  \end{figure}
As evident from the graph figure-\ref{fig:4}, there exist minima of the effective potential which gives stable light rings in the spacetime. This effect is same as observed in \cite{wang2021kerr} 
\\The the effective potential for both magnetized and non-magnetized naked singularities is plotted in the range of radial distance of 0-10.
  \begin{figure}
      \includegraphics[width=95mm,scale=0.2]{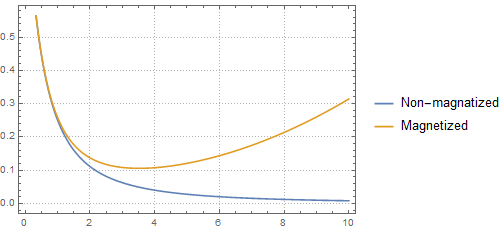}
      \caption{$V_{eff} \longrightarrow r$ of magnetized and non-magnetized naked singularity (M=1; B=0.5; $\theta=\pi/2$ )}
      \label{fig:5}
  \end{figure}
 It is evident from the graph figure-\ref{fig:5} that the magnetic field clearly impacts the effective potential of the null geodesics.There is perfect fitting of the curve at lesser radial distance. The 'magnetized' curve increases exponentially with the radial distance which doesn't happen in the case of the non-magnetized.
 \textbf{This also suggests that the effect of magnetic field would not be observable locally, only at larger distances the field plays any role.}
\begin{figure}
      \includegraphics[width=80mm,scale=0.2]{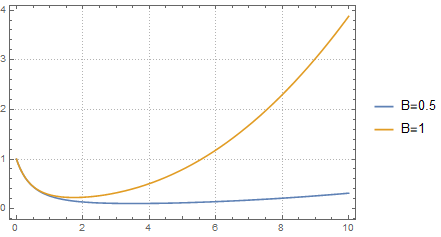}
      \caption{$V_{eff} \longrightarrow r$ of magnetized naked singularity at different external magnetic field (M=1; $\theta=\pi/2$ )}
      \label{fig:6}
  \end{figure}
The graph above figure- \ref{fig:6} is plotted of magnetized naked singularities at different magnetic field parameter B. As the value of B is increased from 0.5 to 1 the behaviour of the graph changes. This shows that, for higher value of magnetic field parameter B, $V_{eff}$ increases rapidly with respect to radial distance than that for the lower value. Ideally, observing the effects at larger distances would not give results of astrophysical significance as the uniform magnetic field is not physically possible throughout the universe. However, these calculations are important in the local frame as the compact object. 
The behaviour of effective potential plots can also be observed with the magnetized and non-magnetized case of the schwarzschild black holes.
\begin{figure}
      \includegraphics[width=95mm,scale=0.2]{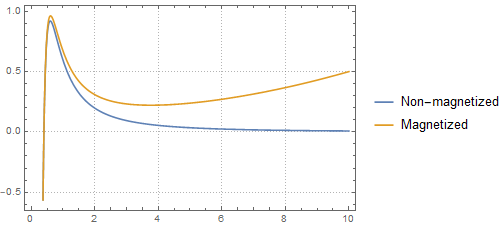}
      \caption{ $V_{eff} \longrightarrow r$ of magnetized and non-magnetized Schwarzschild black hole (M=0.2; B=0.5; $\theta=\pi/2$)}
      \label{fig:7}
  \end{figure}
The graph in figure\ref{fig:7} here is plotted using the equations (24) and (25). Here, the effective potential for for null geodesics would keep on increasing in the magnetized case with respect to the radial distance.
This behaviour is same as the observed in the case of magnetized naked singularity.
\section{Conclusions}
As mentioned in Ref\cite{aliev1989exact}, the Ernst-Wild solutions aimed at magnetizing the black hole metric using a source-
less magnetic field but later on in 1981, it was noted that these were not the solutions to sourceless Enstein-Maxwell system rather it coincides to singular material source which is localized at the polar axis. Hence, these solutions became less meaningful physically. Despite of the limitations, the Ernst technique is a limit to various solutions and can be considered as a key to understanding asymptotic
magnetized solutions before examining other more complex magnetizing techniques.
The effective potential would have an exponential spike as the radial distance increases in the magnetized cases of both, null-like naked singularity and schwarzschild black holes. The shadow image of the magnetized compact object would clearly show the impact of the transformations performed in section II.
\\As shown in Ref\cite{wang2021kerr}, the magnetic field does have a meaningful impact on the shadow of a compact object and is pivotal in understanding real astrophysical situations in the vicinity of the object and as an asymptotic observer.
\textbf{Hence in conclusion this article has summarized the Ernst technique and Harrison transformation, proposed a solution for the null-like naked singularity in Melvin Universe and effective potential plot which suggests the existence of shadow as well and stable light rings.}
\begin{acknowledgments}
I am grateful for the guidance and insights given by Dr. Dipanjan Dey. I am thankful towards P.D. Patel Institute of Applied Science, International Center for Cosmology (ICC) and Charotar University for Science and Technology.
\end{acknowledgments}

\nocite{*}
\bibliography{ref.bib}
\end{document}